\documentclass[aps, prl, twocolumn, groupedaddress, amsmath, amssymb,nofootinbib]{revtex4-1}

\makeatletter

\usepackage{bm,mathrsfs,subfigure}
\usepackage{graphicx}

\renewcommand{\vec}[1]{\bm{#1}} 
\newcommand{\flavor}[1]{\mathsf{#1}}    

\DeclareMathOperator*{\re}{Re}
\DeclareMathOperator*{\im}{Im}
\DeclareMathOperator*{\diag}{diag}

\begin{document}

\title{A new possibility of the fast neutrino-flavor conversion in the pre-shock region of core-collapse supernova}

\author{Taiki Morinaga}
\affiliation{Graduate School of Advanced Science and Engineering, Waseda University, 3-4-1 Okubo, Shinjuku, Tokyo 169-8555, Japan.}

\author{Hiroki Nagakura}
\affiliation{Department of Astrophysical Sciences, Princeton University, Princeton, NJ 08544, USA.}

\author{Chinami Kato}
\affiliation{Department of Aerospace Engineering, Tohoku University, 6-6-01 Aramaki-Aza-Aoba, Aoba-ku, Sendai 980-8579, Japan.}

\author{Shoichi Yamada}
\affiliation{Graduate School of Advanced Science and Engineering, Waseda University, 3-4-1 Okubo, Shinjuku, Tokyo 169-8555, Japan.}

\date{\today}
\begin{abstract}
We make a strong case that the fast neutrino-flavor conversion, one of the collective flavor oscillation modes, commonly occurs in core-collapse supernovae (CCSNe).
It is confirmed in the numerical data obtained in realistic simulations of CCSNe but the argument is much more generic and applicable universally:
the coherent neutrino-nucleus scattering makes the electron lepton number (ELN) change signs at some \textit{inward} direction and trigger the flavor conversion in the \textit{outward} direction in the pre-shock region. 
Although the ELN crossing is tiny and that is why it has eluded recognition so far, it is still large enough to induce the flavor conversion. 
Our findings will have an important observational consequences for CCSNe neutrinos.
\end{abstract}
\maketitle

\noindent\textit{Introduction.}---Neutrinos will give us vital clues not only to the explosion mechanism of core-collapse supernovae (CCSNe) but also to their flavor structures.
In fact, prediction of the luminosities and energy spectra for all neutrino species requires taking into account neutrino oscillations appropriately. 
This is more difficult than previously thought, however, if collective neutrino oscillations occur~\cite{Hannestad2006,Duan2006,Raffelt2007a,Dasgupta2009a,Banerjee2011,Sarikas2012,Sarikas2012a,Cherry2012,Mirizzi2012,Mirizzi2012a,Raffelt2013,Cherry2013,Mirizzi2013,Chakraborty2014,Abbar2015,Dasgupta2015,Yang2017,Tamborra2017,Cirigliano2018,Rrapaj2019,Zaizen:2019ufj}, since they are nonlinear phenomena described with integro-partial differential equations.
No consensus has been reached thus far on whether, when and how the collective oscillation occurs in CCSNe. 
In this \textit{Letter} we make a strong case that the fast neutrino-flavor conversion, one of the collective neutrino oscillation modes, should commonly occurs in the post-bounce phase of CCSNe.

The fast flavor conversion has been extensively studied in the literature~\cite{Chakraborty2016,Sawyer2016,Dasgupta2017a,Izaguirre2017a,Capozzi2017,Dasgupta2018,Abbar2018,Airen2018,Abbar2018a,Capozzi2018a,Abbar2018b,Yi2019,DelfanAzari2019,Martin2019,Shalgar2019}. 
A convenient criterion for its occurrence is supposed to be the ELN crossing, i.e., $\nu_e$ is dominant over $\bar{\nu}_e$ in some propagation directions whereas $\bar{\nu}_e$ overwhelms $\nu_e$ in the other directions. 
\citet{Tamborra2017} searched for such ELN crossings in the numerical data of CCSNe simulations under the assumption of spherical symmetry. 
Paying attention mainly to outward-going neutrinos, they reported negative results. 
More recently, such investigations are extended to the results of multi-dimensional simulations~\cite{Abbar2018a,DelfanAzari2019}. 
\citet{Abbar2018a} found the ELN crossing in some extended domains in the post-shock region. 
On the other hand, \citet{DelfanAzari2019} reported no detection of ELN crossing based on a 2D CCSN model in \citet{Nagakura:2017mnp}.
We stress that these results depend strongly on multi-dimensional effects and may change from model to model.

In this \textit{Letter}, we discuss a new possibility of the fast flavor conversion, based on a more robust argument. We focus on the pre-shock region. 
This is the region ahead of the shock wave, in which cold matter mainly composed of heavy nuclei is imploding toward the shock. 
We argue that the ELN crossing is produced rather commonly by the coherent scattering of neutrinos on these heavy nuclei, with $\bar{\nu}_e$ being scattered more often than $\nu_e$, which sets the stage for the fast flavor conversion.

\citet{Capozzi2018a} pointed out recently that collisional processes are important to generate the fast flavor conversion. What they have in mind in their paper, however, is completely different from what we consider in this \textit{Letter}. 
They studied scatterings that occur in the vicinity of the neutrinosphere whereas we investigate the region at much larger radii; 
the scattering processes are also different. 
\citet{Cherry2012,Cherry2013}, \citet{Cirigliano2018} and \citet{Zaizen:2019ufj} also explored the possible effect of scattering of neutrinos on nucleons in the post-shock region, the so-called neutrino halo. 
Time-independence and spherical symmetry they imposed, however,  obscured the role of the fast flavor conversion unfortunately.

As we shall see below, our argument is quite simple and robust: 
the existence of ELN crossing is demonstrated analytically; it is then vindicated by more realistic CCSN simulations. 
Note that our findings have been overlooked so far probably because the ELN crossing is tiny. 
However, such a tiny crossing is actually large enough for the fast flavor conversion to grow substantially. 
It is also intriguing that the flavor conversion always propagates \textit{outward}, which will hence have an impact on the terrestrial observation of supernova neutrinos.

\noindent\textit{Backward scattering on heavy nuclei.}---Now the main claim of this paper: 
coherent scatterings of neutrinos on heavy nuclei produce the ELN crossing in the pre-shock region which is tiny but still sufficient to induce the fast flavor conversion. 
Interestingly, the conversion propagates \textit{outward} as convective instability~\cite{Briggs1964,Landau1997,Capozzi2017,Yi2019}.
We will substantiate this contention shortly.

The shock wave generated at core bounce is stalled in the core and becomes an accretion shock at $r\sim 200\operatorname{km}$. 
Matter outside this stagnant shock is cold and hence mainly composed of heavy nuclei and is falling almost freely onto the shock front. 
Neutrinos emitted from the neutrinosphere located much deeper inside ($r \lesssim 50\operatorname{km}$) are moving outward almost freely outside the shock, since the matter density is low there. 
A small fraction of these neutrinos are back-scattered by nuclei, however, and produce the inward-going population. 
Since $\bar{\nu}_e$ has higher energies than $\nu_e$ on average while the luminosities are similar between them, the inward-going population is dominated by $\bar{\nu}_e$. 

This can be demonstrated more quantitatively with the so-called bulb model, in which neutrinos are emitted from the neutrino surface half-isotropically. 
For concreteness, we assume that the energy spectra of neutrinos are expressed as $f_{\nu}(E)\propto E^{\alpha_\nu} e^{-(3+\alpha_\nu)E/\bar{E}_{\nu}}$~\cite{Keil2003,Tamborra2012,Mirizzi2012,Supp}.
Hereafter the index $\nu$ represents $\nu_e$ or $\bar\nu_e$.
Without interactions with matter, all neutrinos are going outward, being confined in a cone. 
Their angular distributions are given as \cite{Supp}
\begin{align}
    \mathscr{G}^{\mathrm{bulb}}_{\nu}(\mu) =& 2\operatorname{cm}^{-1}\left(\dfrac{50\operatorname{km}}{R_{\nu}}\right)^2 \left(\dfrac{L_{\nu}}{10^{52}\operatorname{erg/s}}\right) \left(\dfrac{10\operatorname{MeV}}{\bar{E}_{\nu}}\right)\nonumber\\
    &\times \Theta\left(\mu-\sqrt{1-\left(R_\nu/r\right)^2}\right),
    \label{eq:IntensityBulb}
\end{align}
where $R_\nu$ is the radius of the neutrinosphere, $L_\nu$ and $\bar{E}_\nu$ are the luminosity and average energy of neutrino, respectively, $\mu$ is cosine of the zenith angle measured from the local radial direction and $\Theta$ is the step function. 
The ELN angular distribution is given by $\mathscr{G}_{\nu_e}-\mathscr{G}_{\bar\nu_e}$.
It is normally found that the ELN is positive and its intensity is of the order of $\gtrsim 10^{-1}\operatorname{cm}^{-1}$ in the outward direction ($\mu\sim 1$). 

The population of inward-going neutrinos ($\mu \sim -1$) can be estimated from this outward-going population and the matter distribution as follows. 
The density profile outside the shock front is approximately expressed as $\rho(r)\propto r^{-\beta}$ as a function of the radial position $r$. 
The rate of coherent scattering is estimated with the formula given in \citet{Bruenn1985} together with the assumption $A\simeq \mathrm{const.},\ \frac{Z-N}{A}\simeq 0$ for the average mass ($A$), proton ($Z$) and neutron ($N$) numbers of nuclei. 
Then the angular distribution of $\nu$ is derived by line integrations as~\cite{Supp}
\begin{align}
    \mathscr{G}^{\mathrm{scat}}_{\nu}&(\mu) \simeq 2\times 10^{-4}\operatorname{cm}^{-1}\dfrac{4+\alpha_\nu}{(3+\alpha_\nu)(3+\beta)}\left(\frac{A}{56}\right)\nonumber\\
    &\times\left(\dfrac{\rho_{\mathrm{sh}}}{10^7\operatorname{g/cm^3}}\right)\left(\dfrac{R_{\mathrm{sh}}}{200\operatorname{km}}\right)^\beta \left(\dfrac{200\operatorname{km}}{r}\right)^{1+\beta}\nonumber\\
    &\times\left(\dfrac{L_{\nu}}{10^{52}\operatorname{erg/s}}\right)\left(\dfrac{\bar{E}_{\nu}}{10\operatorname{MeV}}\right)\left[(\mu+1) + \frac{1}{4}\left(\frac{R_\nu}{r}\right)^2\right]
    \label{eq:IntensityScattered}
\end{align}
up to the lowest order of $(\mu+1)$ and $(R_\nu/r)$, where $R_{\mathrm{sh}}$ is the shock radius and $\rho_{\mathrm{sh}}$ is the matter density just outside the shock front. 
The leading angular dependence reflects the fact that the coherent scattering is strongly forward-peaked, $\propto (1+ \cos \theta )$, where $\theta$ is the scattering angle~\cite{Bruenn1985}. 
In the limit of $r \to \infty$, the outward-going neutrinos become all radially-going actually and there is no neutrino going radially-inward. 
At finite radii, however, there remains a small finite contribution, giving the second term in the last factor. Note that the difference in $R_\nu$ between $\nu_e$ and $\bar{\nu}_e$ is included only in this term. 
As a result, the ELN ($\mathscr{G}^{\mathrm{scat}}_{\nu_e} - \mathscr{G}^{\mathrm{scat}}_{\bar\nu_e}$) becomes negative as long as $L_\nu \bar{E}_\nu$ is larger for $\bar{\nu}_e$ than for $\nu_e$ at angles that satisfy $1 \gg (1 + \mu) \gg (R_\nu / r)^2/4$. 
The absolute value of ELN is estimated typically to be $\gtrsim 10^{-6} \operatorname{cm}^{-1}$, which will be also vindicated later by realistic simulations.

The different signs of ELN for the outward and inward directions imply that there occurs an ELN crossing in between.
The growth rate of the fast flavor conversion is roughly given by the geometric mean of the ELN intensities at their positive and negative parts (see below)~\cite{Note1}. 
It is estimated to be $\gtrsim 10^{-4}\operatorname{cm}^{-1} = 1/(100\operatorname{m})$, which is large enough for the fast flavor conversion to develope sufficiently in the time scale of CCSNe.

\noindent\textit{Growth rates of flavor conversion.}---Before moving to the realistic numerical models, we give here some mathematical formulae that will be employed there for quantitative analyses. 
The initial phase of the collective neutrino flavor conversion can be studied by the linear stability analysis~\cite{Banerjee2011,Izaguirre2017a,Airen2018}. 
Flavor evolutions are described by the kinetic equations for the density matrices of neutrinos $\flavor{f}$:
\begin{align}
    v\cdot\partial\flavor{f}(x,\Gamma) = -i[\flavor{H}(x,\Gamma),\flavor{f}(x,\Gamma)] + \mathcal{C}[\flavor{f}],
    \label{eq:Liouville}
\end{align}
where $x\equiv(t,\vec{x})$ denotes the position in spacetime, $\Gamma\equiv(E,\vec{v})$ the energy ($E>0$ for neutrino and $E<0$ for antineutrino) and flight direction and $(v^\mu) \equiv (1,\vec{v})$; 
Hamiltonian $\flavor{H}$ is given as $\flavor{H}(x,\Gamma) = \flavor{H}_{\mathrm{vac}}(E)+\flavor{H}_{\mathrm{int}}(x,\vec{v})$ with the vacuum oscillation term $\flavor{H}_{\mathrm{vac}}(E) \equiv \flavor{M}^{2}/2E$ and the potential term $\flavor{H}_{\mathrm{int}}(x,\vec{v}) \equiv v\cdot\flavor{\Lambda}(x)$;  
$\flavor{M}^2$ is the mass-squared matrix and $\flavor{\Lambda}$ is the 4-current of leptons defined as $\flavor{\Lambda}^\mu(x)\equiv \sqrt{2}G_F\left[\diag\left(\{j^\mu_\alpha(x)\}\right) + \int d\Gamma \flavor{f}(x,\Gamma)v^\mu\right]$ with $j^{\mu}$ being the number current of the charged lepton specified by $\alpha$ and $\int d\Gamma \equiv \int_{-\infty}^\infty \frac{dE E^2}{2\pi^2}\int\frac{d^2\vec{v}}{4\pi}$;
$\mathcal{C}$ is the collision term. 

In the region of our current concern, $H_\mathrm{vac}$ is smaller than $H_\mathrm{int}$ and is dropped in the following analysis. 
This implies that only the fast flavor conversion is considered. 
The vacuum-mass term, $H_\mathrm{vac}$, plays the role of an instigator of the flavor conversion in this context, generating initial perturbations. 
If the maximum wave number of vacuum oscillation, $k_{\mathrm{vac}} \equiv \frac{1}{\hbar c}\frac{\Delta m^2_{\mathrm{max}}}{2E} = \frac{10\operatorname{MeV}}{E}\times 6.6 \times 10^{-6}\operatorname{cm^{-1}}$~\cite{PhysRevD.98.030001}, becomes comparable to the growth rate $\sigma$ (see below) of the fast flavor conversion, however, $H_\mathrm{vac}$ should be reinstated and the slow mode needs to be also considered~\cite{Airen2018}. 
The collision term $\mathcal{C}[f]$ is also neglected, since it is important not in the flavor conversion itself but in setting the background for it~\cite{Capozzi2018a}.

We work in the framework of 2-flavor mixing.
Then a small perturbation around the flavor eigenstate is expressed as
\begin{align}
    \flavor{f}(x,\Gamma) = 
    \begin{pmatrix}
        f_{\nu_e}(\Gamma) & 0\\
        0 & f_{\nu_x}(\Gamma)
    \end{pmatrix}
    + \dfrac{f_c(\Gamma)}{2}
    \begin{pmatrix}
        0 & S(x,\Gamma)\\
        \bar{S}(x,\Gamma) & 0
    \end{pmatrix},
\end{align}
where $f_c(\Gamma) \equiv f_{\nu_e}(\Gamma) - f_{\nu_x}(\Gamma)$ and the small off-diagonal component is denoted by $S$. 
Defining further the energy-integrated off-diagonal component  $\mathscr{S}(x,\vec{v}) \equiv e^{i\Lambda_c\cdot x} \int_{-\infty}^{\infty}\frac{dE E^2}{2\pi^2}S(x,\Gamma)$ with $\Lambda_c^\mu \equiv \sqrt{2}G_F\left[j_e^\mu-j_x^\mu + \int d\Gamma f_c(\Gamma)v^\mu\right]$ and the angular intensity of ELN $\mathscr{G}(\vec{v}) \equiv \sqrt{2}G_F\int_{-\infty}^{\infty}\frac{dE E^2}{2\pi^2}f_c(\Gamma)$, we can recast Eq. (\ref{eq:Liouville}) for the off-diagonal component into
\begin{align}
    v\cdot (i\partial) \mathscr{S}(x,\vec{v}) + \int \dfrac{d^2 \vec{v}'}{4\pi} \mathscr{G}(\vec{v}') v\cdot v' \mathscr{S}(x,\vec{v}') = 0
    \label{eq:LinearizedLiouville}
\end{align}
to the linear order of $\mathscr{S}$. 
Note that the variation of $\Lambda$ is neglected, since we consider a patch of space much smaller than the background scale height and a period of time much shorter than the typical hydrodynamical time scale. 
For the plane wave ansatz $\mathscr{S}(x,\vec{v})\equiv Q(\vec{v})e^{ik\cdot x}$, a nontrivial solution of Eq.~(\ref{eq:LinearizedLiouville}) exists iff
\begin{align}
    \det\Pi(k)=0
    \label{eq:DR}
\end{align}
is satisfied for the polarization tensor given as
\begin{align}
    \Pi^{\mu\nu}(k) = \eta^{\mu\nu} + \int \dfrac{d^2\vec{v}}{4\pi} \mathscr{G}(\vec{v})\dfrac{v^\mu v^\nu}{v\cdot k}.
\end{align}

The fast flavor conversion, which is regarded here as instability of the flavor eigenstate, occurs when the solution of Eq.~(\ref{eq:DR}): 
$k^0 = \omega(\vec{k})$ has a positive imaginary part for some $\vec{k}\in\mathbb{R}^3$. 
We normally need to solve Eq.~(\ref{eq:DR}) numerically, not an easy task~\cite{Note2}.

\noindent\textit{Realistic models.}---Below we vindicate the above argument given for the bulb model by quantitatively analyzing the data obtained in our CCSN simulations with the full Boltzmann neutrino transport.
Importantly, the ELN crossings in the pre-shock region are confirmed in many of our models~\cite{Sumiyoshi:2005ri,Nagakura:2019evv} and also in those of Garching group, which are publicly available~\cite{Note3}. 
\citet{Tamborra2017} reported that there was no ELN crossing in the latter models, which is not true, however. 
In the following analysis, we employ a numerical data of a spherically symmetric $11.2M_{\odot}$ CCSN model~\cite{Nagakura:2018qpg} as a representative case.

\begin{figure}[htb]
    \includegraphics[width=\linewidth]{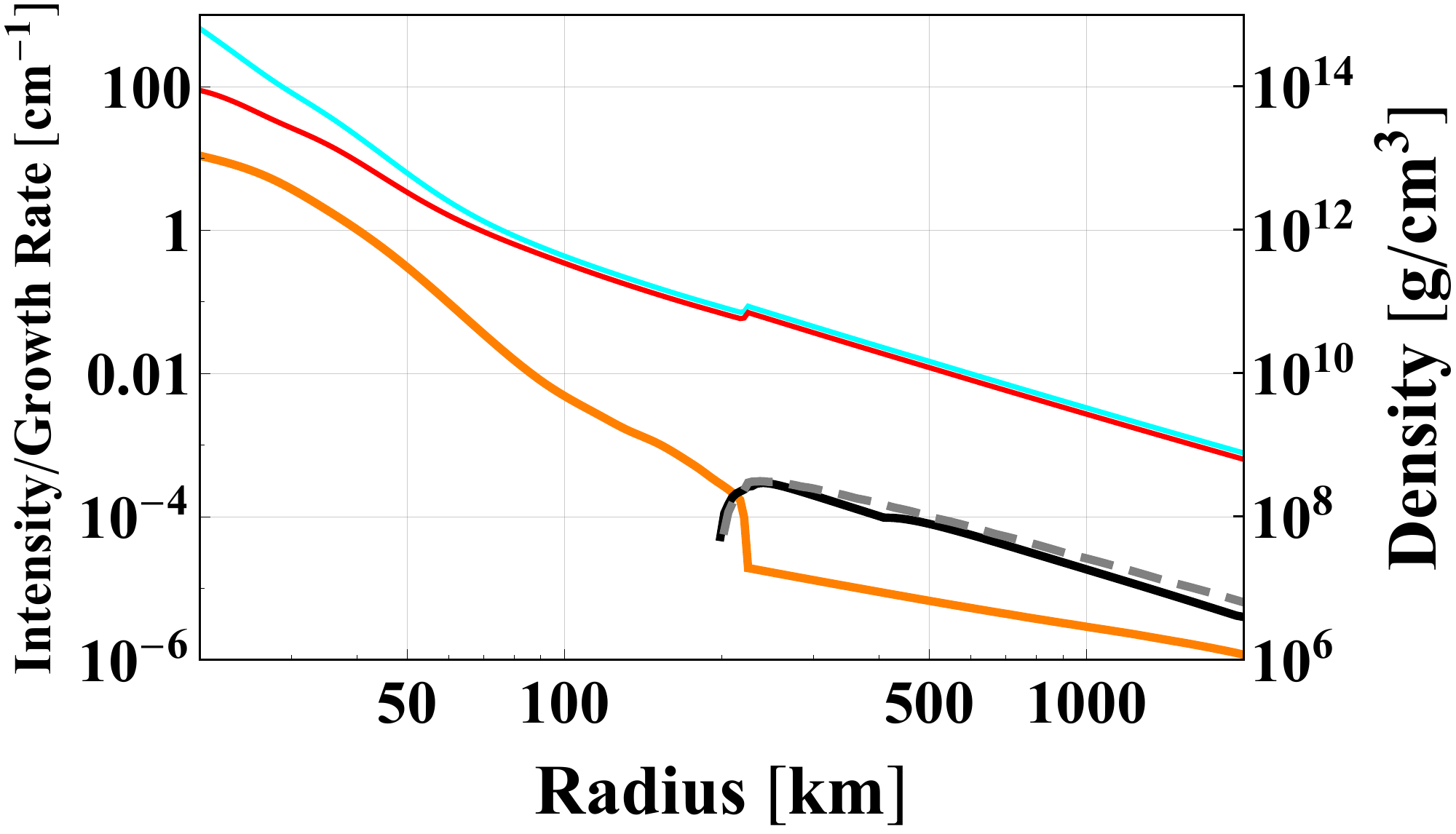}
    \caption{The radial profiles of the baryonic mass density (orange) and the number density of $\nu_e$ (cyan) and $\bar{\nu}_e$ (red) multiplied with $\sqrt{2}G_F(\hbar c)^2$.
     The black solid- and gray dashed lines represent the growth rate of the fast flavor conversion for the standard- ($(N_E,N_\mu)=(20,10)$) and high- ($(N_E,N_\mu)=(30,40)$) resolution simulations, respectively ($N_E$ and $N_\mu$ denote numbers of the energy- and angular grid points, respectively). 
     The time is $100\operatorname{ms}$ after bounce and the shock wave is located at $\sim 223\operatorname{km}$.}
    \label{fig:GrowthRateMap}
\end{figure}

\begin{figure*}[htb]
    \subfigure[]{
        \includegraphics[width=0.31\linewidth]{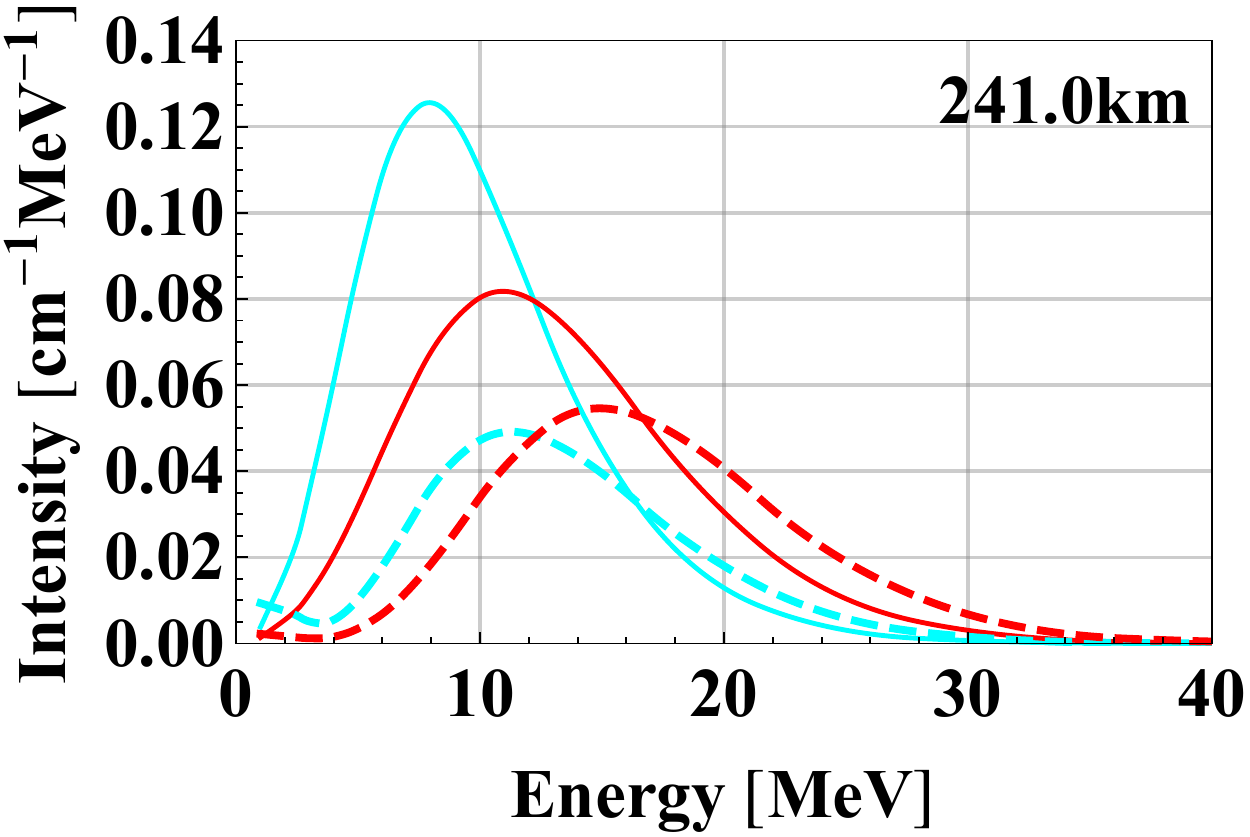}
        \label{fig:EnergySpectra}
    }
    \subfigure[]{
        \includegraphics[width=0.31\linewidth]{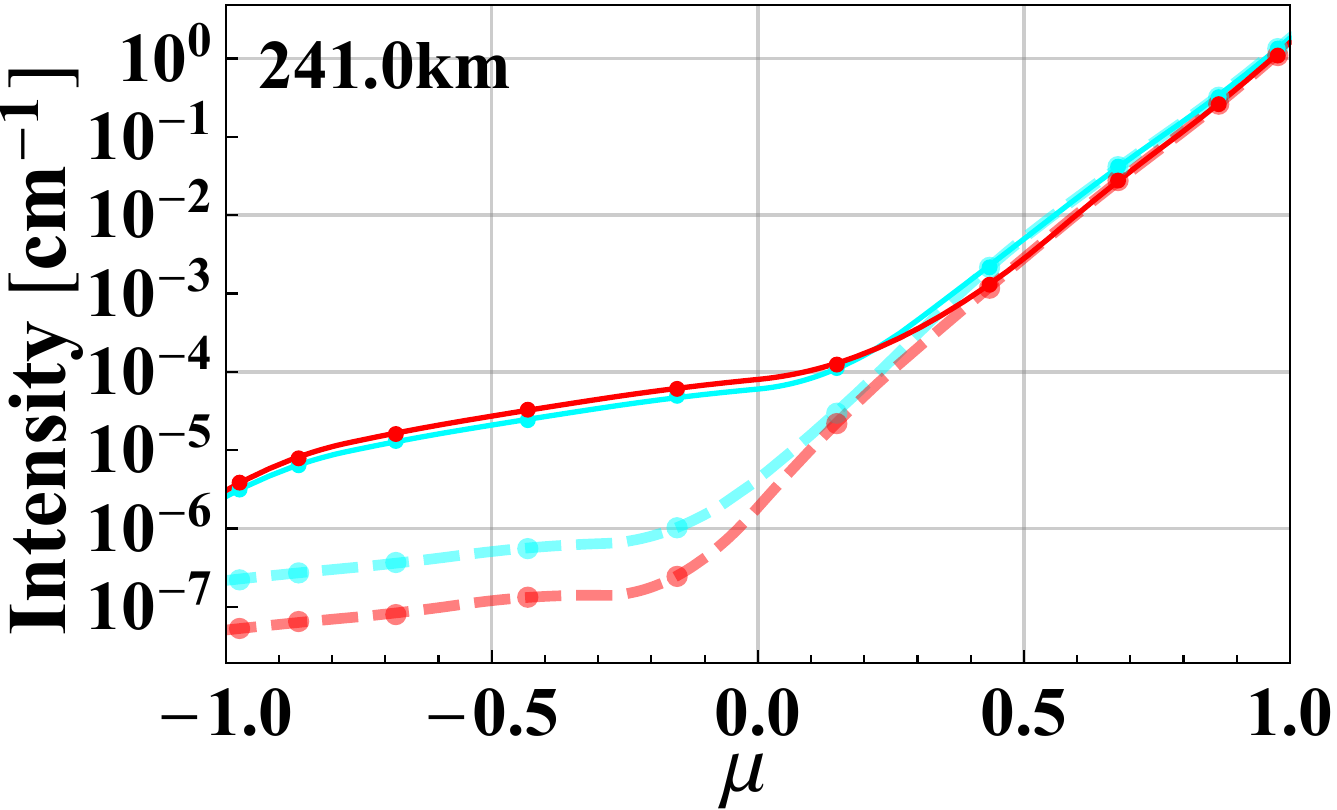}
        \label{fig:AngularDistribution}
    }
    \subfigure[]{
        \includegraphics[width=0.31\linewidth]{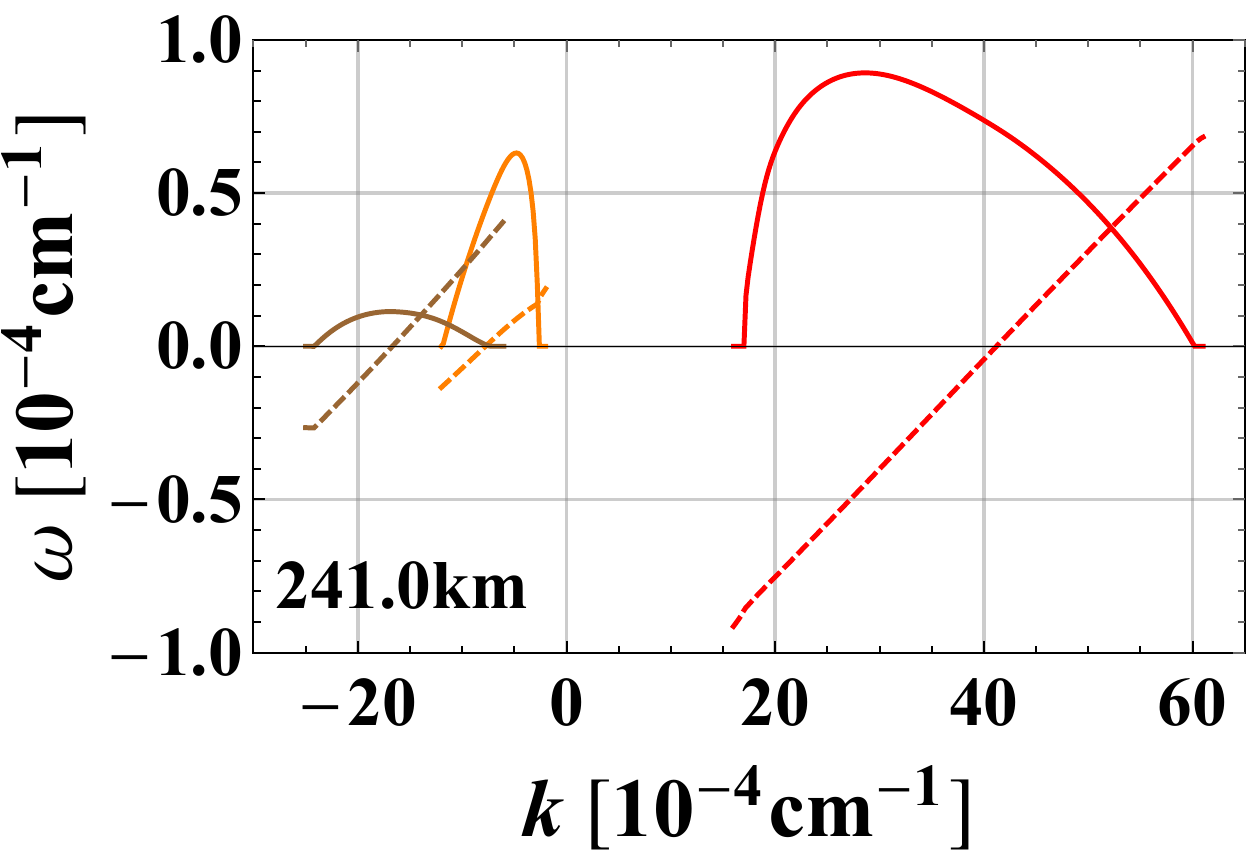}
        \label{fig:DR}
    }

    \caption{\subref{fig:EnergySpectra}: The energy spectra of $\nu_e$ (cyan) and $\bar{\nu}_e$ (red) at $r=241\operatorname{km}$. 
    The solid and dashed lines are for $\mu=0.97$ (outgoing neutrinos) and for $\mu=-0.87$ (ingoing neutrinos), respectively. 
    A factor of $10^5$ is multiplied for the latter. 
    \subref{fig:AngularDistribution}: The angular distributions of neutrinos at the same radius with the same notation for colors. 
    Dashed lines represent the results for the simulation without scatterings of heavy nuclei.
    \subref{fig:DR}: Complex $\omega$ as a function of real $k$ for unstable modes at $r=241\operatorname{km}$ derived by solving Eq. (\ref{eq:DR}). 
    The solid and dashed lines represent $\im\omega$ and $0.05\times\re\omega$, respectively. 
    The time is $100\operatorname{ms}$ after bounce.}
	\label{fig:241km}
\end{figure*}

Figure~\ref{fig:GrowthRateMap} portrays the radial profiles of neutrino number densities and baryonic mass density as well as the approximate estimate of the growth rate of the fast flavor conversion given by the following formula:
\begin{align}
\sigma \sim \sqrt{-\left(\int_{\mathscr{G}(\vec{v})>0}\dfrac{d^2\vec{v}}{4\pi}\mathscr{G}(\vec{v})\right)\left(\int_{\mathscr{G}(\vec{v})<0}\dfrac{d^2\vec{v}}{4\pi}\mathscr{G}(\vec{v})\right)},
\label{eq:GrowthRate}
\end{align}
which is not bad indeed as confirmed later by linear analysis.

As shown in Fig.~\ref{fig:GrowthRateMap}, the fast flavor conversion occurs at the pre-shock region and its growth rate is $\sim 10^{-4}\operatorname{cm}^{-1}$.
It should be stressed that the result is not an artifact by numerical diffusions in our CCSN simulations; 
indeed, the same simulation but with much higher resolutions yields essentially the same results (the gray dashed line in the same figure).
On the other hand, the fast flavor conversion is suppressed in the post-shock region. 
It is attributed to the fact that almost all heavy nuclei are photo-dissociated in the post-shock flows, which substantially reduces scattering opacities.
In addition, the isotropic emission of $\nu_e$ via the electron capture by free protons is enhanced by shock heating and becomes the dominant weak-process for inward-going neutrinos behind the shock wave~\cite{Note4}. 
As a result, the tiny ELN crossing that could be induced by the scattering is washed out and $\nu_e$ dominates over $\bar{\nu}_e$ in all directions.

We turn our attention to the detailed characteristics of the neutrino distributions in momentum space.
For outward-going neutrinos, the average energy, which roughly corresponds to the energy at the peak of the number spectrum, is higher for $\bar\nu_e$ than $\nu_e$, whereas the height of the peak of the spectrum is higher for $\nu_e$ than $\bar\nu_e$ (see solid lines in Fig.~\ref{fig:EnergySpectra}); 
as a result, the number density of $\nu_e$ is slightly larger than that of $\bar\nu_e$, i.e., the ELN is positive (see also the solid lines at $\mu > 0$ in Fig.~\ref{fig:AngularDistribution}).
For inward-going neutrinos, on the other hand, both the height of the peak of the spectrum and the average energy are higher for $\bar\nu_e$ than $\nu_e$ and hence $\bar\nu_e$ is more abundant than $\nu_e$ (see dashed lines in Fig.~\ref{fig:EnergySpectra}), i.e., the ELN is negative. 
This indicates that the ELN crossing occurs, which is exactly what we predicted from our toy model. 
Indeed, it is confirmed that the neutrino angular distributions intersect at $\mu \sim 0.2$ as shown in Fig.~\ref{fig:AngularDistribution}.

To see more clearly the role of the scattering by heavy nuclei, we perform an additional simulation, in which we turn it off. 
The angular distributions of neutrinos obtained in this simulation are displayed as dashed lines in Fig.~\ref{fig:AngularDistribution}. 
It is apparent that the outward-going neutrinos are almost intact whereas the inward-going neutrinos are strongly affected, in which neutrinos are much less abundant and, more importantly, the ELN crossing disappears.
We can hence conclude that the coherent scattering by heavy nuclei plays a crucial role in generating the ELN crossing.

Fig.~\ref{fig:DR} displays the dispersion relation (DR) at $r=241\operatorname{km}$ for $\vec{k}$ parallel to the radial direction, which gives the growth rate of the fast flavor conversion more precisely than Eq.~(\ref{eq:GrowthRate}). 
Note that the maximum growth rate derived from DR is $\sim 10^{-4}\operatorname{cm}^{-1}$, which is roughly the same value estimated by Eq.~(\ref{eq:GrowthRate}) ($2.94\times 10^{-4}\operatorname{cm}^{-1}$).
More interestingly, the group velocity of these unstable modes ($v_{\mathrm{g}}=d\re\omega/dk$) is $\sim 0.7c$ and always positive, which implies that the flavor conversion proceeds in the outward direction.

\begin{figure}[htb]
    \includegraphics[width=\linewidth]{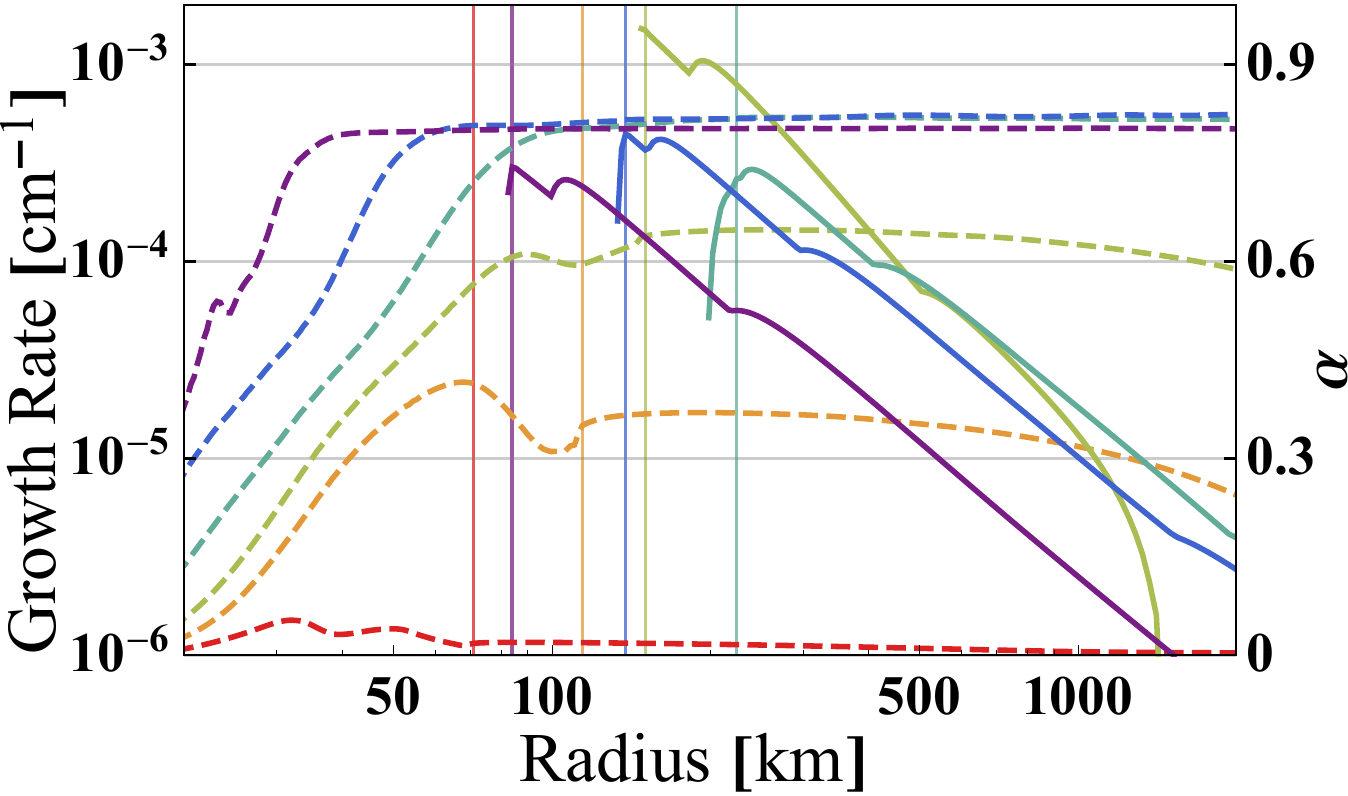}
    \caption{Growth rates of the fast flavor conversion estimated by Eq.~(\ref{eq:GrowthRate}) (solid lines) and the ratio of $n_{\bar{\nu}_e}$ to $n_{\nu_e}$, $\alpha$ (dashed lines) as functions of radius at some different times: 
    $10\operatorname{ms}$ (red), $30\operatorname{ms}$ (orange), $50\operatorname{ms}$ (lime), $100\operatorname{ms}$ (green), $200\operatorname{ms}$ (blue) and $400\operatorname{ms}$ (purple). 
    The thin vertical lines indicate the shock positions at the same times.
    The lack of red and orange solid lines in this figure means no fast flavor conversion at the corresponding times. 
    }
    \label{fig:GrowthRateAsymmetryMap}
\end{figure}

Fig.~\ref{fig:GrowthRateAsymmetryMap} shows the growth rates of the fast flavor conversion as a function of radius at different times. 
One can see that the conversion is suppressed in the early post-bounce phase (up to $\sim 30\operatorname{ms}$ after bounce in this CCSN model). 
This is simply because $\bar\nu_e$ emissions are suppressed at early times.
Once $\bar\nu_e$ is produced substantially (at $\sim 50\operatorname{ms}$), it is confirmed that the ELN crossing occurs in the pre-shock region and is sustained for the rest of the post-bounce phase.

\noindent\textit{Conclusion.}---In this \textit{Letter} we have presented a new possibility of the fast neutrino-flavor conversion in CCSNe. 
We have argued that it should be ubiquitous in the pre-shock region in the post bounce phase except for the very early period ($\lesssim 30\operatorname{ms}$ after bounce). 
The key ingredient is the coherent neutrino-nucleus scattering.
We have demonstrated both analytically and numerically that the scattering induces the ELN crossing and then triggers the fast flavor conversion.
We also found that the group velocities of unstable modes are always positive irrespective of their phase velocities, i.e., the fast flavor conversion should have an influence on the terrestrial observation of supernova neutrinos.

\begin{acknowledgments}
We are grateful to S. Abbar for valuable discussions.
T.M is supported by JSPS Grant-in-Aid for JSPS Fellows (No. 19J21244) from the Ministry of Education, Culture, Sports, Science and Technology (MEXT), Japan. 
H.N was supported by Princeton University through DOE SciDAC4 Grant DE-SC0018297 (subaward 00009650).
Large-scale storage of numerical data is supported by JLDG constructed over SINET4 of NII.
This work is also supported by the Grant-in-Aid for the Scientific Research (Nos. 15K05093, 25870099, 26104006, 16H03986, 17H06357, 17H06365), HPCI Strategic Program of Japanese MEXT and K computer at the RIKEN (Project ID: hpci 160071, 160211, 170230, 170031, 170304, hp180179, hp180111, hp180239) and Waseda University Grant for Special Research Projects (Project number: 2018K-263).
 
\end{acknowledgments}

%

\onecolumngrid
\clearpage
\begin{center}
  \textbf{\large Supplemental Material}
\end{center}
\setcounter{equation}{0}
\setcounter{figure}{0}
\renewcommand{\theequation}{S\arabic{equation}}
\renewcommand{\thefigure}{S\arabic{figure}}

\section{Derivation of Eq. (1)}
\begin{figure}[htb]
    \subfigure[]{
        \includegraphics[width=0.46\linewidth]{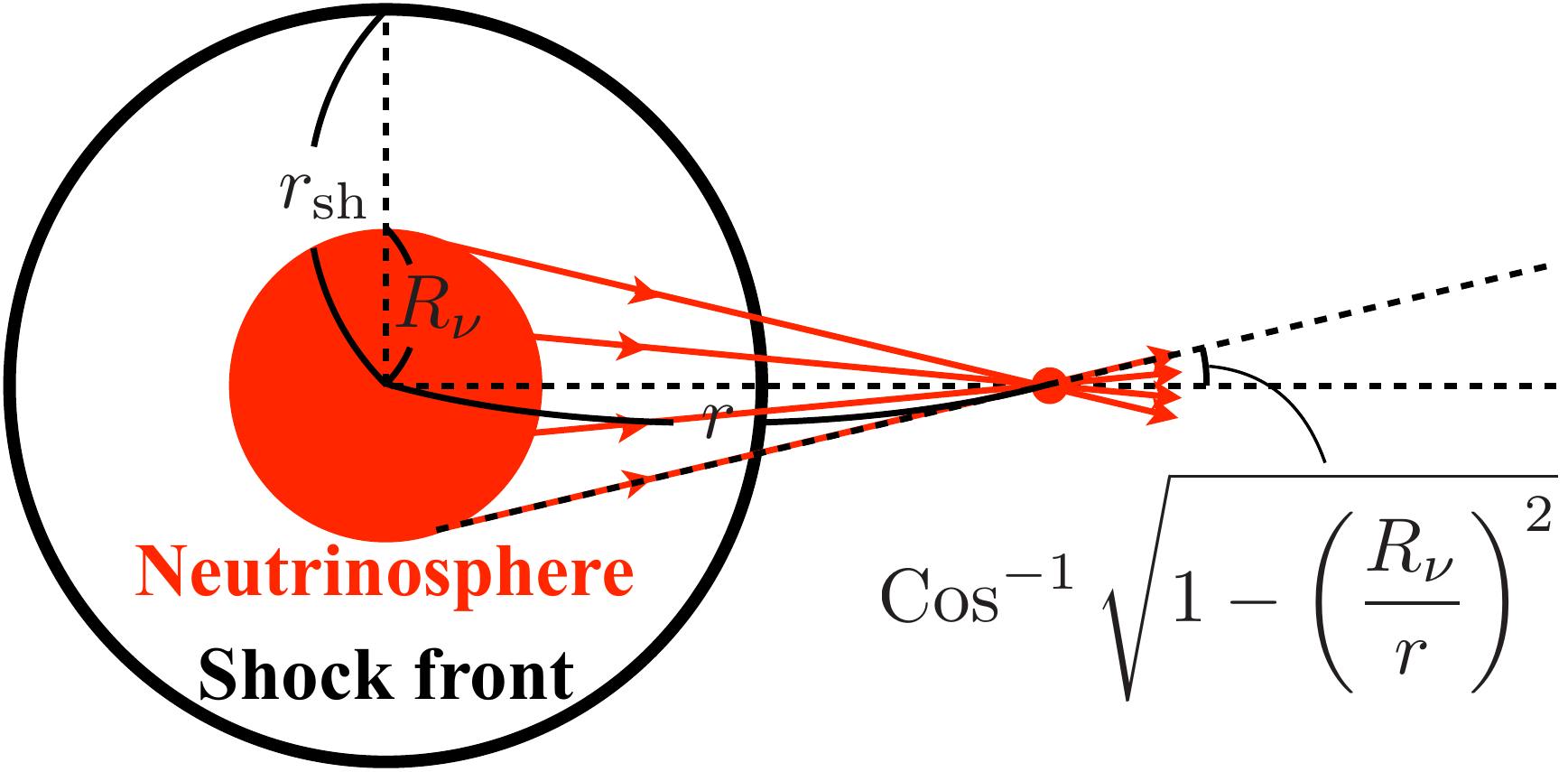}
        \label{fig:BulbModel}
    }
    \subfigure[]{
        \includegraphics[width=0.46\linewidth]{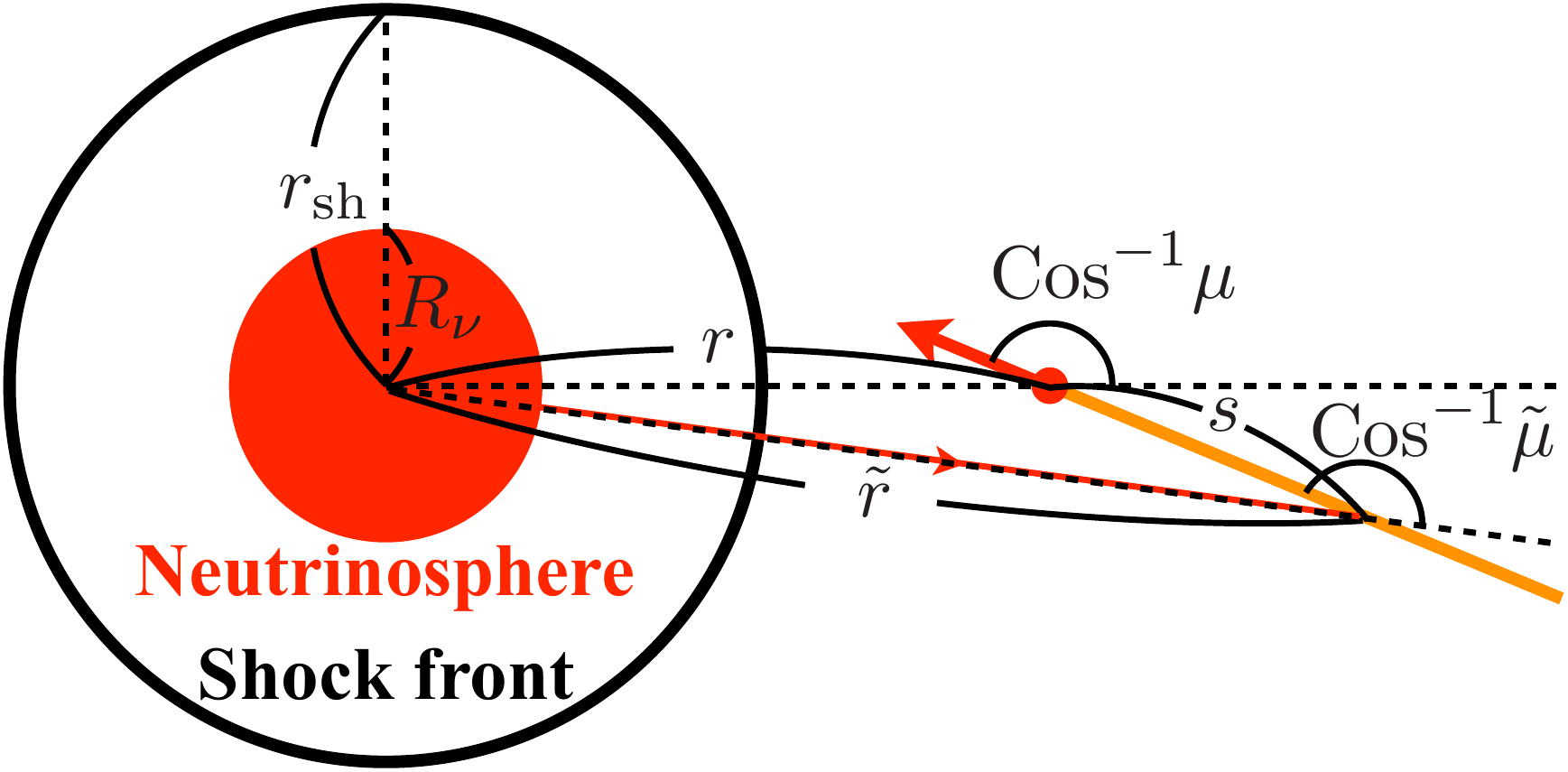}
        \label{fig:Scattering}
    }
    \caption{Schematic pictures of \subref{fig:BulbModel} the neutrino bulb model and \subref{fig:Scattering} coherent scattering backwards. }
    \label{fig:Models}
\end{figure}

We assume that the distribution function of each species of neutrinos takes the following form on the neutrinosphere ($r=R_\nu$):
\begin{align}
    f_{\nu}^{\mathrm{bulb}}(R_{\nu},E,\mu) = CE^{\alpha_\nu} e^{-D E}\Theta(\mu).
\end{align}
In this expression, $\Theta(\mu)$ is the step function and  reflects the assumption in the bulb model that the $\nu$ angular distributions is half-isotropic in the outward direction on the neutrinosphere; 
$C$ and $D$ are constants, which are fixed by the luminosity $L_\nu$ and mean energy $\bar{E}_\nu$ of neutrino from the following relations:
\begin{align}
    L_{\nu} \equiv 4\pi R_{\nu}^2\int dP f_{\nu}^{\mathrm{bulb}}(R_{\nu},E,\mu)Ec\mu =  \frac{4\pi R_{\nu}^2}{(\hbar c)^3}\int_0^\infty \frac{dE E^2}{2\pi^2} \int_{-1}^1\frac{d\mu}{2} f_{\nu}^{\mathrm{bulb}}(R_{\nu},E,\mu)Ec\mu,
\end{align}
\begin{align}
    \bar{E}_\nu \equiv \frac{\int dP f_\nu^{\mathrm{bulb}}(R_\nu,E,\mu)E}{\int dP f_\nu^{\mathrm{bulb}}(R_\nu,E,\mu)},
\end{align}
where we used the notation
\begin{align}
    \int dP \equiv \frac{1}{(\hbar c)^3}\int_{0}^\infty \frac{dE E^2}{2\pi^2}\int\frac{d^2\vec{v}}{4\pi}.
\end{align}
The resultant distribution is given as
\begin{align}
    f_\nu^{\mathrm{bulb}}(R_\nu,E,\mu) = \frac{2\pi(3+\alpha_\nu)^4\hbar^3 c^2 L_\nu}{\Gamma(4+\alpha_\nu)R_\nu^2\bar{E}_\nu^4}\left\{(3+\alpha_\nu)\frac{E}{\bar{E}_\nu}\right\}^{\alpha_\nu} e^{-(3+\alpha_\nu)\frac{E}{\bar{E}_\nu}}\Theta(\mu).
\end{align}
The distribution function at an arbitrary radius $r$ outside the neutrinosphere is then obtained as
\begin{align}
    f_\nu^{\mathrm{bulb}}(r,E,\mu) = \frac{2\pi(3+\alpha_\nu)^4\hbar^3 c^2 L_\nu}{\Gamma(4+\alpha_\nu)R_\nu^2\bar{E}_\nu^4}\left\{(3+\alpha_\nu)\frac{E}{\bar{E}_\nu}\right\}^{\alpha_\nu} e^{-(3+\alpha_\nu)\frac{E}{\bar{E}_\nu}}\Theta\left(\mu-\sqrt{1-\left(\frac{R_\nu}{r}\right)^2}\right).
\end{align}
Note that only the last factor is changed, reflecting the fact that the angular distribution becomes forward-peaked as the radius increases (see Fig.~S\ref{fig:BulbModel}).
Finally the angular intensity is derived from an energy-integration as
\begin{align}
    \mathscr{G}^{\mathrm{bulb}}_\nu =& \frac{\sqrt{2}G_F}{\hbar c}\int_0^\infty \frac{dE E^2}{2\pi^2} f_\nu^{\mathrm{bulb}}(r,E,\mu) \nonumber\\
    =& \frac{\sqrt{2}\hbar^2 cG_F}{\pi}\frac{L_\nu}{R_\nu^2\bar{E}_\nu}\Theta\left(\mu-\sqrt{1-\left(\frac{R_\nu}{r}\right)^2}\right).
    \label{eq:GBulb}
\end{align}

\section{Derivation of Eq. (2)}
Now we evaluate the neutrino population moving inward, which is produced by the coherent scattering of outward-going neutrinos by heavy nuclei. Note that the neutrino angular distribution in the bulb model, which neglects interactions of neutrinos entirely, lacks inward-going neutrinos originally (see Eq.~(\ref{eq:GBulb})). 
In reality, a small fraction of neutrinos are scattered back even outside the neutrinosphere. 
Since this is a minor population particularly outside the shock wave, we can safely ignore their feedback to the major component of outward-going population, say, by re-scatterings.

The neutrinos produced by the coherent scatterings per unit time is obtained just like the collision term of the Boltzmann equation as
\begin{align}
    \left[\frac{\delta f_\nu}{\delta t}\right]_{\mathrm{coll}}(r,E,\mu)= \int \frac{d\vec{v}'}{4\pi}f_\nu(r,E,\mu')R(r,E,\vec{v}',\vec{v})
\end{align}
with the following scattering kernel~\footnote{Ref.~[43] contains a typographical error in the signature of $\cos\theta_{\vec{v}'\vec{v}}$.}:
\begin{align}
    R(r,E,\vec{v}',\vec{v}) = \frac{\hbar^2 c^3 G_F^2}{\pi}n_A(r) A(r)^2\left[\sin^2\theta_W + \frac{1}{2}\frac{Z(r)-N(r)}{A(r)}(1-2\sin^2\theta_W)\right]^2 E^2 (1+\cos\theta_{\vec{v}'\vec{v}})e^{-bE^2(1-\cos\theta_{\vec{v}'\vec{v}})},
\end{align}
where $n_A(r)$, $A(r)$, $Z(r)$ and $N(r)$ are the number density and average mass, proton and neutron numbers of nuclei, respectively, as a function of $r$;
$G_F$ is the Fermi-coupling constant, $\theta_W$ is the Weinberg angle ($\sin^2 \theta_W \simeq 0.231$); $\theta_{\vec{v}'\vec{v}}$ is the angle between $\vec{v}'$ and $\vec{v}$;
the form factor $e^{-bE^2(1-\cos\theta_{\vec{v}\vec{v}'})}$ accounts for the coherency of the scattering and is determined by the ratio of the $\nu$ wavelength to the radius of nuclei. 
Since the distribution function of scattered neutrinos at a given radius $r$ is a sum of all neutrinos scattered at larger radii, it is obtained by line-integration (orange line in Fig.~S\ref{fig:Scattering}) as
\begin{align}
    f^{\mathrm{scat}}_\nu (r,E,\mu) =& \int_0^\infty ds\frac{1}{c}\left[\frac{\delta f_\nu}{\delta t}\right]_{\mathrm{coll}}(\tilde{r},E,\tilde{\mu})\nonumber\\
    =& \frac{\hbar^2 c^2 G_F^2}{\pi} E^2 \int_0^\infty ds\, n_A(\tilde{r}) A(\tilde{r})^2\left[\sin^2\theta_W + \frac{1}{2}\frac{Z(\tilde{r})-N(\tilde{r})}{A(\tilde{r})}(1-2\sin^2\theta_W)\right]^2 \nonumber\\
    & \times \int \frac{d\vec{v}'}{4\pi}(1+\cos\theta_{\vec{v}'\tilde{\vec{v}}})e^{-bE^2(1-\cos\theta_{\vec{v}'\tilde{\vec{v}}})}f_\nu^{\mathrm{bulb}}(\tilde{r},E,\mu'),
\end{align}
where $\tilde{r}$ and $\tilde{\mu}$ are given as follows (see Fig.~\ref{fig:Scattering}):
\begin{align}
    \tilde{r}^2 = r^2 + s^2 - 2rs\mu,
\end{align}
\begin{align}
    \tilde{\mu} = -\frac{s - r\mu}{\sqrt{r^2 + s^2 - 2rs\mu}}.
\end{align}
We assume further that the density profile satisfies a power-law:
\begin{align}
    \rho(r) = \rho_{\mathrm{sh}}\left(\frac{r}{r_{\mathrm{sh}}}\right)^{-\beta}.
\end{align}
In addition, we employ the following approximations:
\begin{align}
    A(r) = \mathrm{const.} (= A),
\end{align}
\begin{align}
    n_A(r) \simeq \frac{\rho(r)}{A m_{a}},
\end{align}
\begin{align}
    \frac{1}{2}\frac{Z(r)-N(r)}{A(r)}(1-2\sin^2\theta_W)\simeq 0,
\end{align}
\begin{align}
    e^{-bE^2(1-\cos\theta_{\vec{v}'\tilde{\vec{v}}})} \simeq 1,
    \label{eq:FormFactor}
\end{align}
where $m_a$ is the atomic mass unit. 
These are all reasonable approximations outside the shock wave in the post-bounce phase.
Then $f^{\mathrm{scat}}_\nu (r,E,\mu)$ is expressed as
\begin{align}
    f^{\mathrm{scat}}_\nu (r,E,\mu) \simeq& \frac{2\hbar^5 c^4 G_F^2 \sin^4\theta_W}{m_a}\frac{(3+\alpha_\nu)^2}{\Gamma(4+\alpha_\nu)} \frac{AL_\nu\rho_{\mathrm{sh}}}{R_\nu^2 \bar{E}_\nu^2}\left(\frac{r_{\mathrm{sh}}}{R_\nu}\right)^\beta\left\{(3+\alpha_\nu)\frac{E}{\bar{E}_\nu}\right\}^{2+\alpha_\nu} e^{-(3+\alpha_\nu)\frac{E}{\bar{E}_\nu}}\nonumber\\
    & \times\int_0^\infty ds \left(\frac{R_\nu}{\tilde{r}}\right)^\beta\int\frac{d\vec{v}'}{4\pi}(1+\cos\theta_{\vec{v}'\tilde{\vec{v}}})\Theta\left(\mu'-\sqrt{1-\left(\frac{R_\nu}{\tilde{r}}\right)^2}\right).
    \label{eq:FScat}
\end{align}

We now evaluate the integrals by expanding $\tilde{\mu}$ and $\left(R_\nu/\tilde{r}\right)^\beta$ in terms of $(\mu+1)$ as
\begin{align}
    \tilde{\mu} = \left\{-1 + \frac{1}{u+1}(\mu+1)\right\}\sum_{n=0}^{\infty}
    \begin{pmatrix}
        -\frac{1}{2}\\
        n
    \end{pmatrix}
    \left\{\frac{-2u}{(u+1)^2}(\mu+1)\right\}^n
\end{align}
and
\begin{align}
    \left(\frac{R_\nu}{\tilde{r}}\right)^\beta = \left(\frac{R}{r}\right)^\beta \frac{1}{(u+1)^\beta}\sum_{n=0}^{\infty}
    \begin{pmatrix}
        -\frac{\beta}{2}\\
        n
    \end{pmatrix}
    \left\{\frac{-2u}{(u+1)^2}(\mu+1)\right\}^n,
\end{align}
respectively. 
In the above equations we define
\begin{align}
    u \equiv \frac{s}{r}.
\end{align}
The second line of Eq.~(\ref{eq:FScat}) is then evaluated as follows:
\begin{align}
    & \int_0^\infty ds \left(\frac{R_\nu}{\tilde{r}}\right)^\beta\int\frac{d\vec{v}'}{4\pi}(1+\cos\theta_{\vec{v}'\tilde{\vec{v}}})\Theta\left(\mu'-\sqrt{1-\left(\frac{R_\nu}{\tilde{r}}\right)^2}\right) \nonumber\\
    =& \frac{1}{2}\int_0^\infty ds \left(\frac{R_\nu}{\tilde{r}}\right)^\beta\left[1-\sqrt{1-\left(\frac{R_\nu}{\tilde{r}}\right)^2} + \frac{\tilde{\mu}}{2}\left(\frac{R_\nu}{\tilde{r}}\right)^2\right] \nonumber\\
    =& \frac{1}{2}\left(\frac{R_\nu}{r}\right)^{2+\beta}r\int_0^\infty du \left[\frac{1}{2(u+1)^{4+\beta}}(\mu+1) + \frac{1}{8(u+1)^{4+\beta}}\left(\frac{R_\nu}{r}\right)^2\right] + \mathrm{h.o.}\nonumber\\
    =& \frac{1}{4(3+\beta)}\left(\frac{R_\nu}{r}\right)^{2+\beta}r\left[(\mu+1) + \frac{1}{4}\left(\frac{R_\nu}{r}\right)^2\right] + \mathrm{h.o.},
\end{align}
where $\mathrm{h.o.}$ means higher order terms in $(\mu+1)$ or $\left(R_\nu/r\right)$. 
The distribution function of the scattered neutrinos $f^{\mathrm{scat}}_\nu (r,E,\mu)$ is finally given as
\begin{align}
    f^{\mathrm{scat}}_\nu (r,E,\mu) \simeq& \frac{\hbar^5 c^4 G_F^2 \sin^4\theta_W}{2 m_a} \frac{(3+\alpha_\nu)^2}{\Gamma(4+\alpha_\nu)(3+\beta)}\frac{AL_\nu\rho_{\mathrm{sh}}}{\bar{E}_\nu^2}\left(\frac{r_{\mathrm{sh}}}{r}\right)^\beta\frac{1}{r} \left\{(3+\alpha_\nu)\frac{E}{\bar{E}_\nu}\right\}^{2+\alpha_\nu} e^{-(3+\alpha_\nu)\frac{E}{\bar{E}_\nu}}\nonumber\\
    & \times \left[(\mu+1) + \frac{1}{4}\left(\frac{R_\nu}{r}\right)^2\right] + \mathrm{h.o.}
\end{align}
The corresponding angular intensity $\mathscr{G}_\nu^{\mathrm{scat}}(r,\mu)$ is derived as
\begin{align}
    \mathscr{G}_\nu^{\mathrm{scat}}(r,\mu) =& \frac{\sqrt{2}G_F}{\hbar c}\int_0^\infty \frac{dE E^2}{2\pi^2} f_\nu^{\mathrm{scat}}(r,E,\mu) \nonumber\\
    \simeq& \frac{\sqrt{2}\hbar^4 c^3 G_F^3 \sin^4\theta_W}{4\pi^2m_a}\frac{4+\alpha_\nu}{(3+\alpha_\nu)(3+\beta)} AL_\nu\bar{E}_\nu \rho_{\mathrm{sh}}\left(\frac{r_{\mathrm{sh}}}{r}\right)^\beta\frac{1}{r}\left[(\mu+1) + \frac{1}{4}\left(\frac{R_\nu}{r}\right)^2\right] + \mathrm{h.o.}
\end{align}

\end{document}